\documentclass[aps,PRB,reprint,superscriptaddress,preprintnumbers]{revtex4-2}
\usepackage[T1]{fontenc}
\usepackage{times}
\usepackage{graphicx,color}
\usepackage{amsfonts,amsmath,amssymb,amsbsy}
\usepackage[colorlinks=true, linkcolor=blue, citecolor=blue, urlcolor=blue]{hyperref}
\usepackage{float}
\usepackage{lettrine}

\let\boldsymbol=\mathbf
\def\emph#1{\textcolor{red}{#1}}

\def\section#1{\medskip\noindent\textbf{#1}\par}
\begin{document}

\title{Nonreciprocal dynamics of ferrimagnetic bimerons}

\author{Laichuan Shen}
\affiliation{School of Science and Engineering, The Chinese University of Hong Kong, Shenzhen, Guangdong 518172, China}

\author{Jing Xia}
\affiliation{College of Physics and Electronic Engineering, Sichuan Normal University, Chengdu 610068, China}

\author{Zehan Chen}
\affiliation{Department of Electronic and Computer Engineering, The Hong Kong University of Science and Technology, Clear Water Bay, Kowloon, Hong Kong, China}
\affiliation{Department of Physics, The Hong Kong University of Science and Technology, Clear Water Bay, Kowloon, Hong Kong, China}

\author{Xiaoguang Li}
\affiliation{Center for Advanced Material Diagnostic Technology, College of Engineering Physics, Shenzhen Technology University, Shenzhen 518118, China}

\author{Xichao Zhang}
\affiliation{Department of Electrical and Computer Engineering, Shinshu University, 4-17-1 Wakasato, Nagano 380-8553, Japan}

\author{\\ Oleg A. Tretiakov}
\affiliation{School of Physics, The University of New South Wales, Sydney 2052, Australia}

\author{Qiming Shao}
\email[Email:~]{eeqshao@ust.hk}
\affiliation{Department of Electronic and Computer Engineering, The Hong Kong University of Science and Technology, Clear Water Bay, Kowloon, Hong Kong, China}
\affiliation{Department of Physics, The Hong Kong University of Science and Technology, Clear Water Bay, Kowloon, Hong Kong, China}
\affiliation{Guangdong-Hong Kong-Macao Joint Laboratory for Intelligent Micro-Nano Optoelectronic Technology, The Hong Kong University of Science and Technology, Hong Kong, China}

\author{Guoping Zhao}
\affiliation{College of Physics and Electronic Engineering, Sichuan Normal University, Chengdu 610068, China}

\author{Xiaoxi Liu}
\affiliation{Department of Electrical and Computer Engineering, Shinshu University, 4-17-1 Wakasato, Nagano 380-8553, Japan}

\author{Motohiko Ezawa}
\email[Email:~]{ezawa@ap.t.u-tokyo.ac.jp}
\affiliation{Department of Applied Physics, The University of Tokyo, 7-3-1 Hongo, Tokyo 113-8656, Japan}

\author{Yan Zhou}
\email[Email:~]{zhouyan@cuhk.edu.cn}
\affiliation{School of Science and Engineering, The Chinese University of Hong Kong, Shenzhen, Guangdong 518172, China}

\clearpage
\begin{abstract}
\noindent
Magnetic bimerons are topologically nontrivial spin textures in in-plane easy-axis magnets, which can be used as particle-like information carriers.
Here, we report a theoretical study on the nonreciprocal dynamics of asymmetrical ferrimagnetic (FiM) bimerons induced by spin currents.
The FiM bimerons have the ability to move at a speed of kilometers per second and do not show the skyrmion Hall effect at the angular momentum compensation point. 
Our micromagnetic simulations and analytical results demonstrate that spin currents are able to induce the nonreciprocal transport and a drift motion of the FiM bimeron even if the system is at the angular momentum compensation point.
By analyzing the current-induced effective fields, we find that the nonreciprocal transport is attributed to the asymmetry of the bimeron structure. 
Our results are useful for understanding the physics of bimerons in ferrimagnets and may provide guidelines for building bimeron-based spintronic devices.
\end{abstract}

\date{\today}

\maketitle

\clearpage


\textit{Introduction.}~Reciprocity is a fundamental principle in many fields, such as in mechanics and thermodynamics~\cite{Takashima_PRB2018}. 
However, when a certain symmetry of the system is broken, the reciprocal relation may be violated and nonreciprocal phenomena appear~\cite{Takashima_PRB2018,Tokura_NC2018,Fan_NanoLett2019}.
Nonreciprocal transport which plays an important role in various application devices, such as in the diode~\cite{Li_RMP2012,Whyte_NC2015,Xing_JAP2016,Song_JMMM2021} and shift register~\cite{Franken_NatNano2012}, has been reported for (quasi-)particles, for instance, electrons~\cite{Ideue_NatPhys2017}, phonons~\cite{Li_RMP2012}, photons~\cite{Toyoda_PRL2015} and magnons~\cite{Tateno_PRAppl2020}.
For topological solitons, e.g. skyrmions~\cite{Roszler_NATURE2006}, they also show such a transport in the asymmetrical racetrack which requires sophisticated reprocessing~\cite{Zhao_Nanoscale2020,Jung_PRB2021,Fook_SciRep2016,Wang_APL2020}. 
However, nonreciprocal transport attributed to the intrinsic characteristics of topological solitons still remains to be discovered.

Different types of topological spin textures have been investigated for a few decades, such as domain walls~\cite{Parkin_Sci2008,Gomonay_PRL2016}, skyrmions~\cite{Muhlbauer_Sci2009,Roszler_NATURE2006,Gobel_PhysRep2021} and bimerons~\cite{Ezawa_PRB2011,Lin_PRB2015,Leonov_PRB2017,Kharkov_PRL2017,Yu_Nat2018,Gobel_PRB2019,Murooka2019,Shen_PRL2020,Zhang_Bimeron2020,Li_npj2020,Nagase_NC2021}, 
which emerge in ferromagnetic (FM)~\cite{Parkin_Sci2008,Muhlbauer_Sci2009,Yu_Nat2018}, ferrimagnetic (FiM)~\cite{Woo_NATCOM2018,Kim_NatMat2017,Caretta_NatNano2018,Xu_PRM2021} and antiferromagnetic (AFM)~\cite{Gomonay_PRL2016,Gao_Nature2020,Jani_Nature2021} materials.
In particular, FM skyrmions are promising as nonvolatile information carriers to serve the future memory and logic computing devices~\cite{Nagaosa_NNANO2013,Finocchio_JPD2016,Fert_NATREVMAT2017,ES_JAP2018,Zhou_NSR2018,Zhang_JPCM2020,Sampaio_NatNano2013}.
However, the FM skyrmion shows the transverse drift during its motion due to the existence of a nonzero Magnus force, which may lead to the annihilation of the fast-moving skyrmion at the sample edge.
This phenomenon is referred to as the skyrmion Hall effect~\cite{Jiang_NatPhys2017,Litzius_NatPhys2017,Zang_PRL2011}.
Compared to the FM skyrmion, the AFM skyrmion is free from the skyrmion Hall effect, as the compensated lattice structures of antiferromagnets lead to a perfect cancellation of the Magnus force~\cite{Zhang_SREP2016A,Barker_PRL2016}. 
However, the compensated magnetic moments in antiferromagnets give rise to the difficulties in detecting AFM spin textures~\cite{Potkina_JAP2020}.
Recently, FiM materials have received great attention, since the AFM spin dynamics is realized in ferrimagnets at the angular momentum compensation point~\cite{Kim_NatMat2017,Hirata_NatNano2019} and unlike the antiferromagnet, even for compensated ferrimagnet, we can detect the magnetization of one sublattice using magnetotransport measurements, such as anomalous Hall effect or tunnel magnetoresistance. 
On the other hand, a magnetic bimeron consisting of two merons is considered as the topological counterpart of a magnetic skyrmion in in-plane magnets and is stabilized in various magnetic materials~\cite{Kharkov_PRL2017,Zhang_Bimeron2020,Lin_PRB2015,Murooka2019,Shen_PRL2020,Li_npj2020,Gobel_PRB2019,Nagase_NC2021}.
Recent reports show that two-dimensional CrCl$_{3}$~\cite{Augustin_NC2021,Lu_NC2020} and van der Waals LaCl/In$_{2}$Se$_{3}$ heterostructures~\cite{Sun_NC2020} are promising candidates for hosting bimerons.
Additionally, the bimeron is a stable solution in ferromagnets~\cite{Yu_Nat2018,Gobel_PRB2019,Shen_PRB2020}, antiferromagnets~\cite{Shen_PRL2020,Li_npj2020,Jani_Nature2021} and frustrated magnets~\cite{Zhang_Bimeron2020,Zhang_APL2021}.
Although a bimeron is topologically equivalent to a skyrmion, the former has richer dynamics~\cite{Li_npj2020,Shen_PRB2020}.

In this work, based on the Landau-Lifshitz-Gilbert (LLG) equation~\cite{Gilbert_IEEE2004}, we theoretically study the current-induced dynamics of FiM bimerons with intrinsic asymmetrical shape.
Numerical and analytical results demonstrate that the FiM bimeron driven by opposite currents could exhibit different speeds, that is, it shows the nonreciprocal dynamics. 
By analyzing the current-induced effective fields, it is found that such a nonreciprocal behavior is attributed to the asymmetry of the bimeron structure.
In addition to the nonreciprocal dynamics, a drift motion of the FiM bimeron may be induced by the current even if the system is at the angular momentum compensation point.

\begin{figure*}[t]
\centering
\centerline{\includegraphics[width=0.70\textwidth]{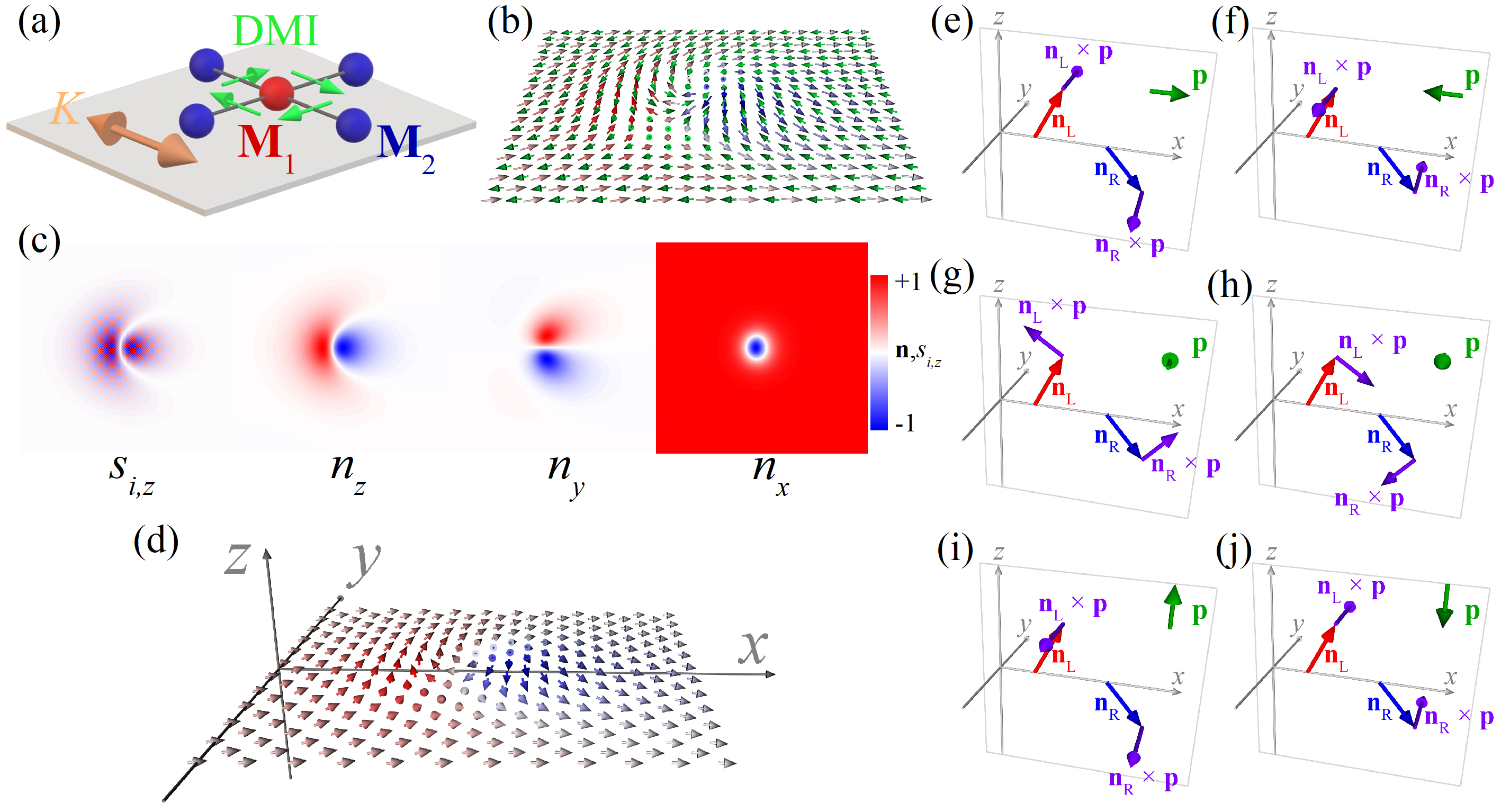}}
\caption{(a) Schematic of the studied model, where $\boldsymbol{M_{\text{1}}}$ and $\boldsymbol{M_{\text{2}}}$ denote two sublattice magnetization of ferrimagnet. The interface-induced DMI and in-plane easy-axis magnetic anisotropy $K$ are considered.
(b) The real-space spin structure of a FiM bimeron.
(c) The components of reduced magnetization $\boldsymbol{s}$ and the N{\'e}el vector $\boldsymbol{n}$ for a FiM bimeron with a positive topological charge $Q$.
(d) The N{\'e}el vector $\boldsymbol{n}$ in real space. The color represents the $z$ components of $\boldsymbol{n}$, and $n_{z}$ of two merons has opposite signs.
(e)-(j) The sketch of two vectors $\boldsymbol{n}_{\text{L}}$ and $\boldsymbol{n}_{\text{R}}$, and the cross product of the vector $\boldsymbol{n}_{\text{L,R}}$ and polarization vector $\boldsymbol{p}$.
} 
\label{FIG1}
\end{figure*}

\textit{Proposal of nonreciprocal transport of FiM bimerons.}~We consider a FiM film with two sublattice magnetization $\boldsymbol{M_{\text{1}}}$ and $\boldsymbol{M_{\text{2}}}$ [Fig.~\ref{FIG1}(a)], and the interfacial Dzyaloshinskii-Moriya interaction (DMI)~\cite{Nagaosa_NNANO2013,Fert_NATREVMAT2017} is introduced, which can be induced at the magnetic layer/heavy metal interface.
To form FiM bimerons, the ferrimagnets with in-plane magnetic anisotropy (such as DyCo$_{5}$~\cite{Unal_PRAppl2016}) are promising materials.   
Here we focus on the study of a FiM film with in-plane easy-axis anisotropy, in which the asymmetrical bimeron is a stable solution, similar to the cases of FM~\cite{Shen_PRB2020} and AFM~\cite{Li_npj2020} bimerons.
For the interfacial DMI shown in Fig.~\ref{FIG1}(a), it is usually responsible for stabilizing the skyrmion with rotational symmetry (the axis of rotation is parallel to the polar axis that is perpendicular to the $x$-$y$ plane)~\cite{Leonov_JPCM2016}.
For the FiM system we consider, the in-plane magnetic anisotropy forces the magnetization to tilt away from the polar axis, which violates the rotational symmetry dominated by DMI.
As reported in Refs.~\cite{Leonov_JPCM2016,Murooka2019}, in the tilted magnetic phases (the magnetization of a homogeneously magnetized state is tilted away from the polar axis), the rotationally symmetrical spin texture is an incompatible form and the asymmetrical spin textures appear.
Figure~\ref{FIG1}(b) shows the spin structure of a FiM bimeron, and the components of its reduced magnetization $\boldsymbol{s}_{i} = \boldsymbol{M}_{i}/\left| \boldsymbol{M}_{i} \right|$ and N{\'e}el vector $\boldsymbol{n} = (\boldsymbol{s}_{i}-\boldsymbol{s}_{j})/2$ are presented in the Fig.~\ref{FIG1}(c).
The N{\'e}el vector $\boldsymbol{n}$ in real space for a FiM bimeron is plotted in Fig.~\ref{FIG1}(d), showing that although the size of the left meron is different from that of the right meron, the bimeron's spin structure still has mirror symmetry about the $x$-$z$ plane.

Additionally, we derive a closed equation for the N{\'e}el vector $\boldsymbol{n}$~\cite{Hals_PRL2011,Kim_PRB2017} (see Supplemental Material~\cite{Shen_SM} for details), $\frac{\mu_{0}\rho^{2}(1+\alpha^{2})} {2\lambda}\ddot{\boldsymbol{n}}\times\boldsymbol{n}+\sigma\dot{\boldsymbol{n}}=-\boldsymbol{n}\times\boldsymbol{f}_{\text{n}}^{*}+\alpha\rho\boldsymbol{n}\times\dot{\boldsymbol{n}}+2\epsilon\boldsymbol{n}\times\boldsymbol{p}\times\boldsymbol{n}$.
$\rho=\frac{M_{\text{S1}}} {\gamma_{1}} + \frac{M_{\text{S2}}} {\gamma_{2}}$ and $\sigma=\frac{M_{\text{S1}}} {\gamma_{1}} - \frac{M_{\text{S2}}} {\gamma_{2}}$ are the staggered and net spin densities, respectively~\cite{Kim_PRB2017}.
$M_{\text{S}i}$, $\gamma_{i}$, $\mu_{0}$, $\lambda$ and $\alpha$ are the saturation magnetization, gyromagnetic ratio, vacuum permeability constant, homogeneous exchange constant and damping constant, respectively.
$\boldsymbol{f}_{\text{n}}^{*}$ and $\epsilon$ relate to the effective field and current density, respectively (Supplemental Material~\cite{Shen_SM}).
In the above equation, only the damping-like spin torque is considered, while the field-like spin torque is not included. The effect of field-like spin torque on the FiM bimeron has been discussed in Supplemental Material~\cite{Shen_SM}.
The above equation indicates that the current-induced effective field relates to the cross product of the N{\'e}el vector $\boldsymbol{n}$ and polarization vector $\boldsymbol{p}$.
Such current-induced effective fields are of interest to us in the following symmetry analysis.
As mentioned earlier, the bimeron's spin structure is symmetric about the $x$-$z$ plane, so that the N{\'e}el vector $\boldsymbol{n}$ is canceled in the $y$ direction and the $y$ component of $\boldsymbol{n}$ will not contribute to the nonreciprocal dynamics.
Thus, we only need to pay attention to the components of $\boldsymbol{n}$ in the $x$-$z$ plane and their corresponding current-induced effective fields ($\boldsymbol{n} \times \boldsymbol{p}$).
As shown in Figs.~\ref{FIG1}(e)-\ref{FIG1}(j), we sketch two vectors $\boldsymbol{n}_{\text{L}}$ and $\boldsymbol{n}_{\text{R}}$ to represent the $x$-$z$-plane components of the N{\'e}el vector $\boldsymbol{n}$ for two merons, respectively.
Based on the symmetry consideration, $\boldsymbol{n}_{\text{L}}$ and $\boldsymbol{n}_{\text{R}}$ are symmetric about the $x$ axis for a bimeron with a symmetrical shape (see Fig. S1 of Supplemental Material~\cite{Shen_SM}), while for the bimeron studied here it has an asymmetrical shape, resulting in the breaking of this symmetry.
Figure~\ref{FIG1}(e) shows the results of $\boldsymbol{n} \times \boldsymbol{p}$ for $\boldsymbol{p} = \boldsymbol{e}_{x}$, where the cross product operation causes the current-induced effective fields to be perpendicular to the $x$-$z$ plane.
When we change the sign of the current, which is equivalent to changing the direction of $\boldsymbol{p}$, \textit{i.e.}, $\boldsymbol{p} = \boldsymbol{e}_{x} \to \boldsymbol{-e}_{x}$, the corresponding current-induced effective fields are still perpendicular to the $x$-$z$ plane [Fig.~\ref{FIG1}(f)].
By comparing Fig.~\ref{FIG1}(e) with Fig.~\ref{FIG1}(f), we see that for $\boldsymbol{p} = \boldsymbol{e}_{x}$ and $\boldsymbol{-e}_{x}$, the current-induced effective fields are symmetric about the $x$-$z$ plane, so that the bimeron does not exhibit the nonreciprocal motion behavior.
Similar mirror symmetry is observed for the case of $\boldsymbol{p} = \pm\boldsymbol{e}_{z}$ [Figs.~\ref{FIG1}(i) and~\ref{FIG1}(j)], so there is no nonreciprocal phenomenon.
	
However, for $\boldsymbol{p} = \boldsymbol{e}_{y}$ [Fig.~\ref{FIG1}(g)] and $\boldsymbol{-e}_{y}$ [Fig.~\ref{FIG1}(h)], $\boldsymbol{n} \times \boldsymbol{p}$ is in the $x$-$z$ plane, and obviously $\boldsymbol{n}_{\text{L}} \times \boldsymbol{e}_{y}$ is not mirror-symmetrical to $\boldsymbol{n}_{\text{L}} \times (\boldsymbol{-e}_{y})$, so that the opposite currents have different effects on the meron on the left in Fig.~\ref{FIG1}(d) [this result also applies to the meron on the right in Fig.~\ref{FIG1}(d)].
For a bimeron with a symmetrical shape, as mentioned above, $\boldsymbol{n}_{\text{L}}$ and $\boldsymbol{n}_{\text{R}}$ are symmetric about the $x$ axis, indicating that the effect of the positive current on the left meron (the right meron) is equivalent to that of the negative current on the right meron (the left meron).
Therefore, although opposite currents have different effects on each meron, the structural symmetry causes the opposite currents to have the same effects on the whole, so that the bimeron with a symmetrical shape will not show nonreciprocal transport, which has been confirmed in Supplemental Material~\cite{Shen_SM}.
For the FiM bimeron studied in this work, it has an asymmetrical shape [Fig.~\ref{FIG1}(c)], resulting in the presence of nonreciprocal phenomena.
Namely, nonreciprocal transport is attributed to the asymmetry of the bimeron structure.

\begin{figure*}[t]
\centering
\centerline{\includegraphics[width=0.85\textwidth]{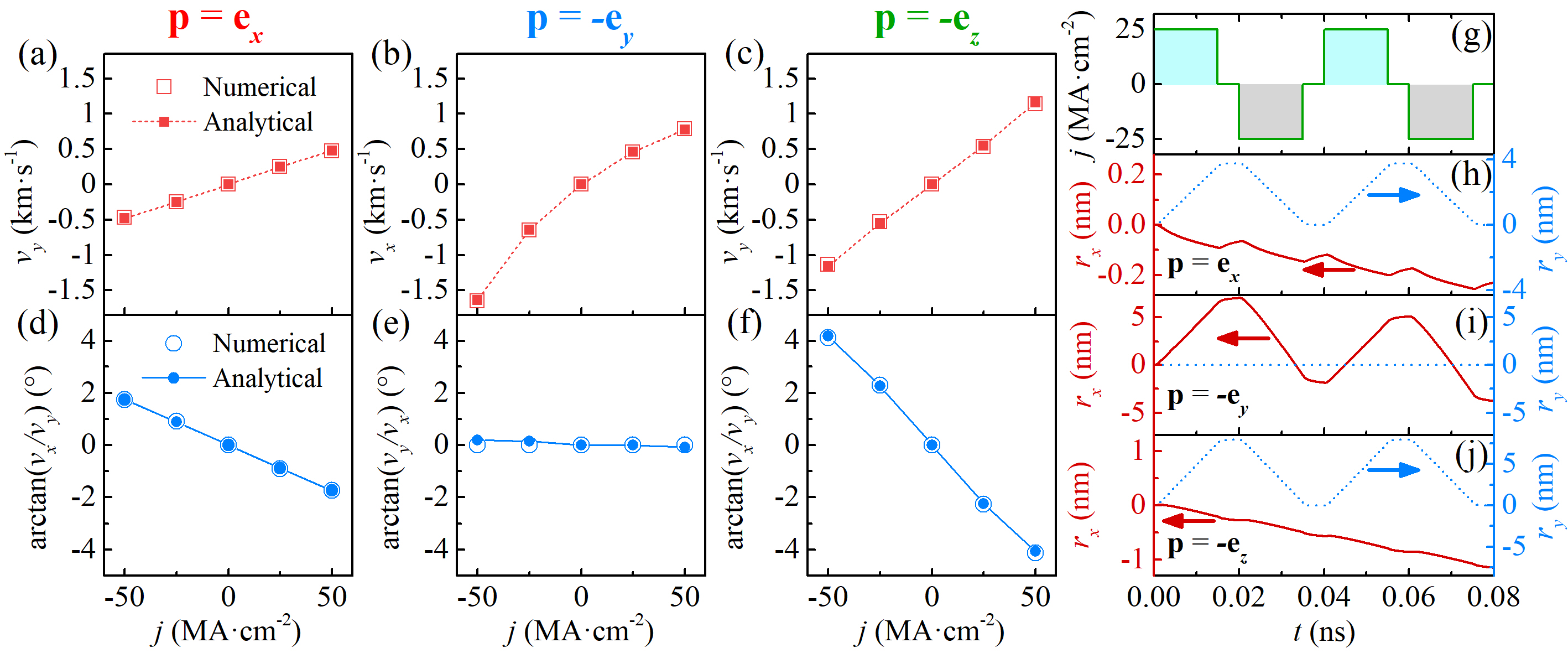}}
\caption{(a)-(c) The FiM bimeron speeds as functions of the current density $j$ for $\boldsymbol{p} = \boldsymbol{e}_{x}$, $\boldsymbol{-e}_{y}$, and $\boldsymbol{-e}_{z}$.
(d)-(f) The angle between the actual and desired motion directions for $\boldsymbol{p} = \boldsymbol{e}_{x}$, $\boldsymbol{-e}_{y}$, and $\boldsymbol{-e}_{z}$.
Here, the FiM bimeron has a positive $Q$, and the numerical and analytical results are obtained by solving LLG equation and Eq.~(\ref{eq:1}), respectively.
(g) The alternating current pulse adopted in our simulation.
(h)-(j) The time evolution of the guiding center ($r_{x}$, $r_{y}$) for different polarization vectors $\boldsymbol{p}$.
$\alpha = 0.05$, $M_{\text{S}1} = 1.1M_{\text{S}2}$ and $\gamma_{1}=1.1\gamma_{2}$.}
\label{FIG2}
\end{figure*}

\textit{Current-induced nonreciprocal transport and drift motion of FiM bimerons.}~To verify the above analysis, we have simulated the magnetization dynamics of FiM bimerons and obtained the bimeron speeds, as shown in Figs.~\ref{FIG2}(a)-\ref{FIG2}(c).
We indeed observe that only when the polarization vector $\boldsymbol{p}$ is along the $y$ direction (perpendicular to the symmetry plane of the bimeron's spin structure), the FiM bimeron driven by opposite currents has different speeds, that is, it exhibits the nonreciprocal transport [Fig.~\ref{FIG2}(b)].
As expected by the above symmetry analysis, for the cases where $\boldsymbol{p}$ is along the $x$ or $z$ directions [Figs.~\ref{FIG2}(a) and~\ref{FIG2}(c)], such a nonreciprocal transport does not appear.  
Here we employ the LLG equation with the damping-like spin torque~\cite{Slonczewski_JMMM1996,Tomasello_SciRep2014} to simulate the dynamics of FiM bimerons, and the simulation details are given in Supplemental Material~\cite{Shen_SM}.   
We also simulate the creation of FiM bimerons (see Figs. S4 and S5 of Supplemental Material~\cite{Shen_SM}) and the creation process is given in Supplemental Movie 1-3.
Moreover, based on the definition of the guiding center $r_{i}=-1/(4\pi Q) \int{[i\boldsymbol{n}\cdot(\partial_{x}\boldsymbol{n}\times\partial_{y}\boldsymbol{n})]}\text{d}S$~\cite{Komineas_PRB2015},
we obtain the time evolution of $r_{i}$ and the bimeron velocity $v_{i}=\dot{r}_{i}$ (see Fig. S8 of Supplemental Material~\cite{Shen_SM}).
$Q = -1/(4\pi)\int{\text{d}S}[\boldsymbol{n}\cdot(\partial_{x}\boldsymbol{n}\times\partial_{y}\boldsymbol{n})]$ is the topological charge~\cite{Barker_PRL2016,Lin_PRB2015}.

We now discuss the current-induced drift motion of FiM bimerons.
Figures~\ref{FIG2}(a)-\ref{FIG2}(c) present the bimeron speeds in the desired motion direction (it is in $y$, $x$ and $y$ directions for $\boldsymbol{p} = \boldsymbol{e}_{x}$, $\boldsymbol{-e}_{y}$ and $\boldsymbol{-e}_{z}$, respectively). 
The FiM bimeron at the angular momentum compensation point has an ability to move with a speed of about km s$^{-1}$, similar to the AFM spin textures~\cite{Shen_PRL2020}.
However, spin currents may induce a drift speed which is perpendicular to the desired motion direction, even if the FiM system is at the angular momentum compensation point (\textit{i.e.}, $\sigma=\frac{M_{\text{S1}}} {\gamma_{1}} - \frac{M_{\text{S2}}} {\gamma_{2}} = 0$). 
As shown in Figs.~\ref{FIG2}(d) and \ref{FIG2}(f) where $\boldsymbol{p} = \boldsymbol{e}_{x}$ and $\boldsymbol{-e}_{z}$ respectively, the angle between the actual and desired motion directions is not zero, that is, the FiM bimeron shows a drift motion, and such an angle increases with the applied currents $j$. For the case of $\boldsymbol{p} = \boldsymbol{-e}_{y}$, the drift motion can be safely disregarded [Fig.~\ref{FIG2}(e)].  

To explain the simulation results, we derived the Thiele equation (see Supplemental Material~\cite{Shen_SM} for details)~\cite{Thiele_PRL1973,Kim_PRB2017,Tveten_PRL2013,Tretiakov_PRL2008}, 
from which we obtain the steady motion speeds,
\begin{equation}
\begin{pmatrix} v_{x}\\v_{y} \end{pmatrix}=\frac{1} {\eta} \begin{pmatrix} \alpha L_{yy} & -G-\alpha L_{xy}\\G-\alpha L_{xy} & \alpha L_{xx} \end{pmatrix} \begin{pmatrix} F_{x}\\F_{y} \end{pmatrix},\tag{1}
\label{eq:1}
\end{equation}
where $\eta=\alpha^{2}(L_{xx}L_{yy}-L_{xy}^{2})+G^{2}$.
$L_{ij} = \mu_{0}\rho t_{z} \int{\text{d}S}(\partial_{i}\boldsymbol{n}\cdot\partial_{j}\boldsymbol{n})$ and $G = 4\pi Q \mu_{0} t_{z} \sigma$ with the layer thickness $t_{z}$.
$F_{i}$ denotes the driving force induced by the damping-like spin torque (its expression is given in Supplemental Material~\cite{Shen_SM}).
If $G = 0$, Eq.~(\ref{eq:1}) indicates that the bimeron speed is inversely proportional to the damping (see Fig. S9 of Supplemental Material~\cite{Shen_SM}).

\begin{figure}[t]
\centerline{\includegraphics[width=0.40\textwidth]{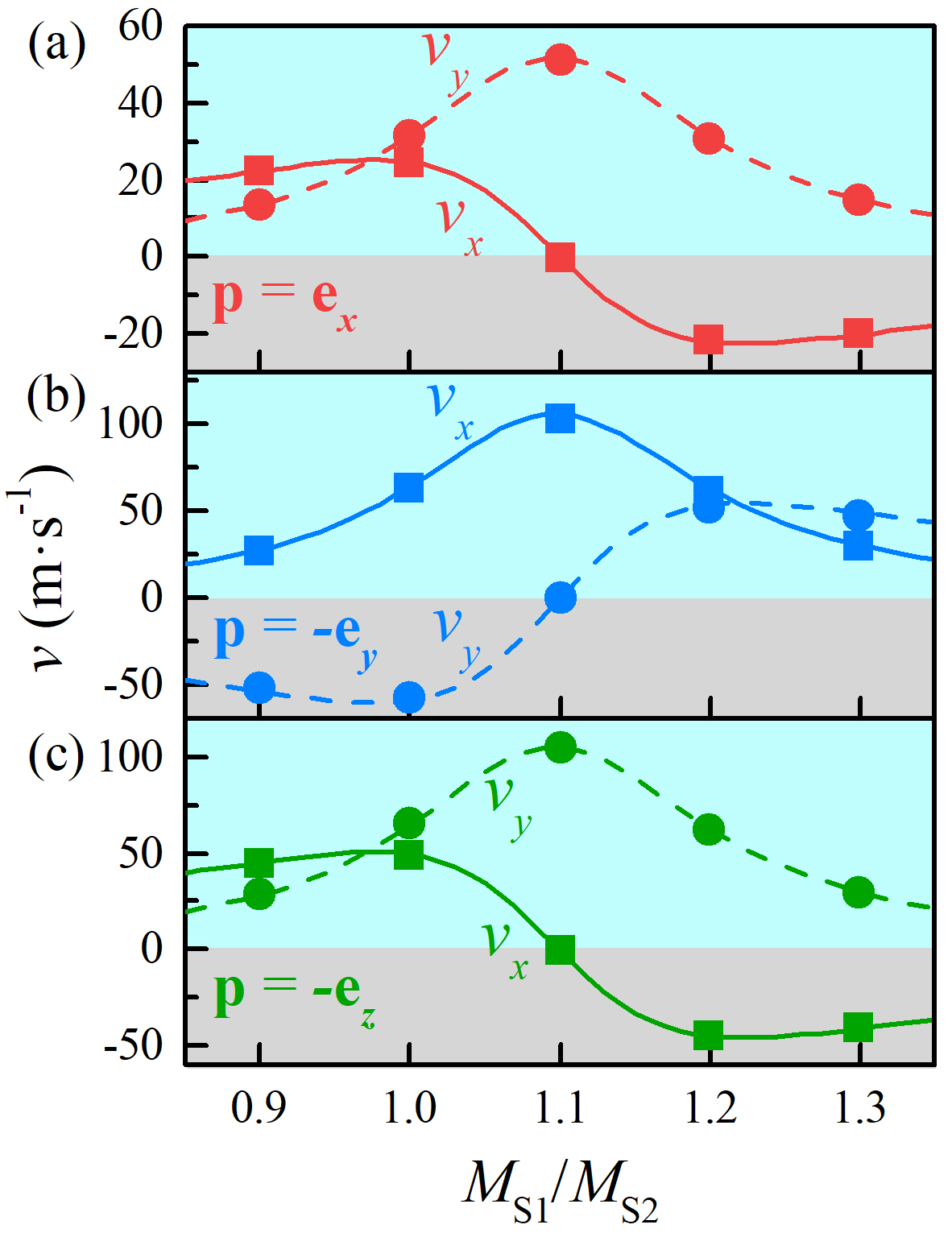}}
\caption{The numerical (symbols) and analytical (curves) velocities for a FiM bimeron in systems with different values of $M_{\text{S}1}/M_{\text{S}2}$, where three polarization vectors (a) $\boldsymbol{p} = \boldsymbol{e}_{x}$, (b) $\boldsymbol{p} = \boldsymbol{-e}_{y}$ and (c) $\boldsymbol{p} = \boldsymbol{-e}_{z}$ are considered.
Here, $j = 5$ MA cm$^{-2}$, $M_{\text{S}2} = 376$ kA m$^{-1}$, $\gamma_{1}=1.1\gamma_{2}$, and other parameters are the same as those used in Fig.~\ref{FIG2}.
The numerical and analytical results are obtained from the LLG equation and Eq.~(\ref{eq:1}), respectively.}
\label{FIG3}
\end{figure}

To verify the above analytical formula, we simulate the motion of FiM bimerons and calculate the bimeron velocities for different values of $M_{\text{S}1}/M_{\text{S}2}$. Figure~\ref{FIG3} shows the comparison of the numerical and analytical velocities, where the analytical velocities for all polarization vectors $\boldsymbol{p}$ are in good agreement with the numerical results. Moreover, Fig.~\ref{FIG3} shows that one of the velocity components ($v_{x}$, $v_{y}$) is symmetric about $M_{\text{S}1}/M_{\text{S}2} = 1.1$, while the other component is antisymmetric. From Eq.~(\ref{eq:1}), we see that one of the velocity components is proportional to $1/(C+G^{2})$ with a constant $C$, \textit{i.e.}, $v_{i} \propto 1/(C+G^{2})$, while the other component $v_{j} \propto G/(C+G^{2})$, where $G \propto (M_{\text{S}1}/M_{\text{S}2} - \gamma_{1}/\gamma_{2})$ because we fixed the values of $M_{\text{S}2}$ and $\gamma_{1}$ in the simulation. For $v_{i} \propto 1/(C+G^{2})$ it presents a symmetric curve, while for $v_{j} \propto G/(C+G^{2})$, an antisymmetric curve is obtained.

Assuming that the main driving force is in the $y$ direction ($F_{y} \neq 0$) and the system is at the angular momentum compensation point ($G = 0$), from Eq.~(\ref{eq:1}), the drift speed is obtained, $v_{x} = \alpha (L_{yy} F_{x} - L_{xy} F_{y})/\eta$.
Note that $L_{yy}$ (or $L_{xx}$) is always nonzero for spin textures.
According to the above formula of the drift speed, we find that there are two factors which cause the drift motion even if $G = 0$. 
The first factor is the presence of a nonzero $L_{xy}$~\cite{Murooka2019} and the second factor is that an additional force perpendicular to the desired motion direction is induced by the applied currents.
In order to verify the above analysis, we calculated the numerical values of $\boldsymbol{L}$ and $\boldsymbol{F}$ (Figs. S10 and S12 of Supplemental Material~\cite{Shen_SM}), and then substituting them into Eq.~(\ref{eq:1}) gives the analytical drift speeds which are consistent with the numerical results, as shown in Figs.~\ref{FIG2}(d)-\ref{FIG2}(f).
Moreover, the numerical values of $\boldsymbol{L}$ and $\boldsymbol{F}$ confirm that the drift motion presented in Figs.~\ref{FIG2}(d) and \ref{FIG2}(f) is due to the presence of a nonzero $L_{xy}$ and an additional force [they originate from the deformation of the bimeron's spin structure after the spin currents are applied (Fig. S13 of Supplemental Material~\cite{Shen_SM})].
The drift speed due to the nonzero $L_{xy}$ is greater than that due to the presence of an additional force for the case of Fig.~\ref{FIG2}(d), while the drift speed in Fig.~\ref{FIG2}(f) is dominated by the additional force.  

Since FiM bimerons exhibit the nonreciprocal transport, an alternating current pulse presented in Fig.~\ref{FIG2}(g) induces the bimeron to show a ratchet motion~\cite{Reichardt_arXiv,Bellizotti_PRB2021,Gobel_SciRep2021,Ma_PRB2017,Reichhardt_ARCMP2017} if we take $\boldsymbol{p} = \boldsymbol{-e}_{y}$ [Fig.~\ref{FIG2}(i)].
Thus, FiM bimerons are ideal information carriers in AC racetrack storage devices~\cite{Gobel_SciRep2021}.
For $\boldsymbol{p} = \boldsymbol{e}_{x}$ and $\boldsymbol{-e}_{z}$, the bimeron does not show the nonreciprocal motion in the $y$ direction and the final value of $r_{y}$ is zero [Figs.~\ref{FIG2}(h) and~\ref{FIG2}(j)], while due to the presence of the drift motion [Figs.~\ref{FIG2}(d) and~\ref{FIG2}(f)], the final values of $r_{x}$ are not equal to zero.

\begin{figure}[t]
\centerline{\includegraphics[width=0.45\textwidth]{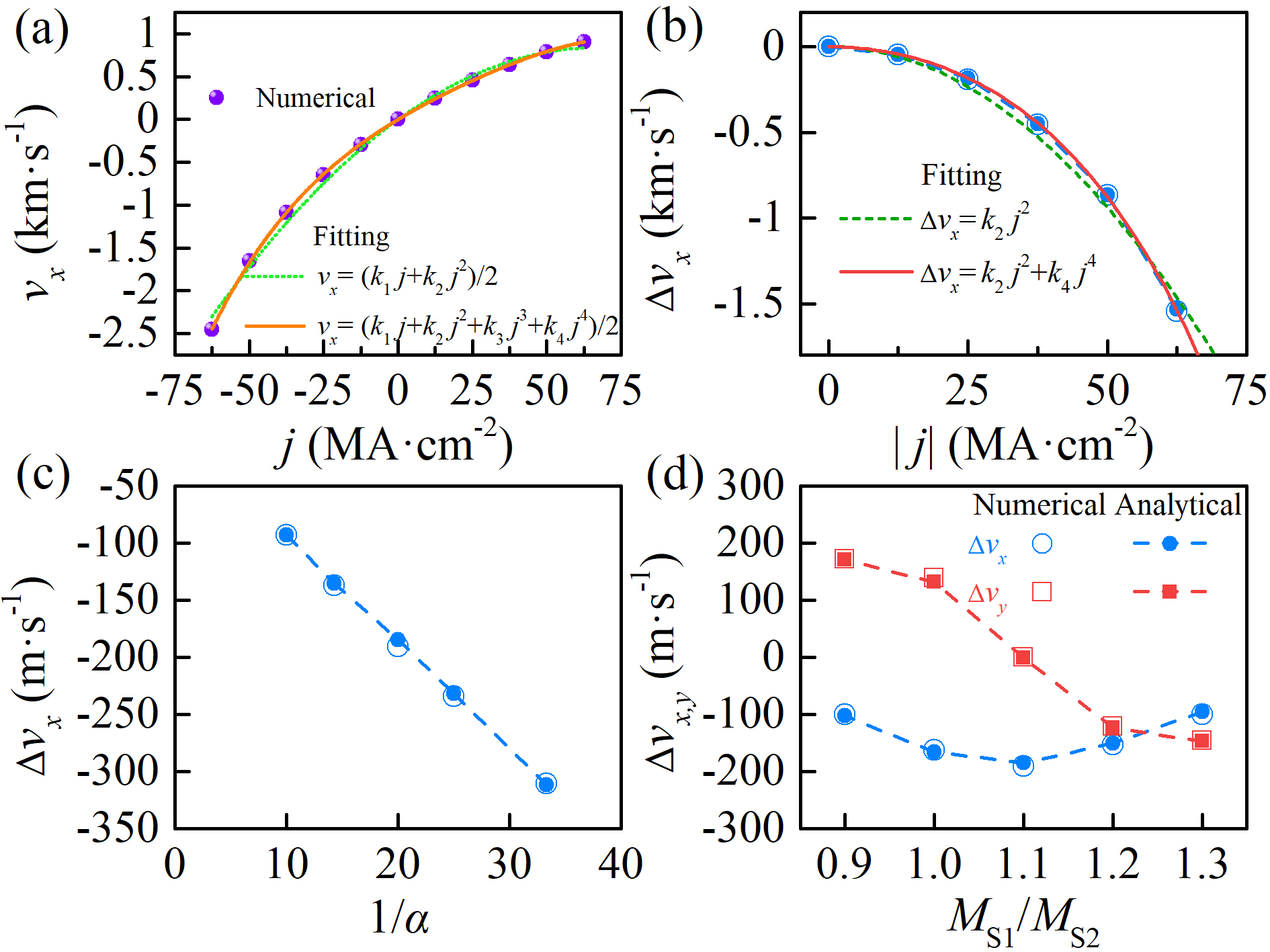}}
\caption{(a) The FiM bimeron speeds $v_{x}$ as functions of the current density $j$. The symbols are obtained from the numerical simulations and the curves are the fitting results.
For the case where the fitting function is $v_{x}=\sum_{l=1}^{2}k_{l}j^{l}/2$, $k_{1} = 50.158$ and $k_{2} = -0.375$ [the unit of $k_{l}$ is m s$^{-1}$ (MA cm$^{-2}$)$^{-l}$]. For $v_{x}=\sum_{l=1}^{4}k_{l}j^{l}/2$, $k_{1} = 41.726$, $k_{2} = -0.274$, $k_{3} = 3.04 \times 10^{-
3}$ and $k_{4} = -3.085 \times 10^{-5}$.
The adopted parameters are the same as those used in Fig.~\ref{FIG2}(b).
(b)-(d) The speed differences $\Delta v_{i}$ as functions of the current density $j$, the damping constant $\alpha$ and the ratio of $M_{\text{S}1}$ and $M_{\text{S}2}$. 
The numerical and analytical results are obtained by solving LLG equation and Eq.~(\ref{eq:1}), respectively.
In panel (b), the green dashed and red solid curves are the fitting results of $\Delta v_{x}=k_{2}j^{2}$ (with $k_{2} = -0.375$) and $\Delta v_{x}=k_{2}j^{2}+k_{4}j^{4}$ (with $k_{2} = -0.274$ and $k_{4} = -3.085 \times 10^{-5}$), respectively.  
The default parameters are $\boldsymbol{p} = \boldsymbol{-e}_{y}$, $j = \pm 25$ MA cm$^{-2}$, $\alpha = 0.05$, $M_{\text{S}1} = 1.1M_{\text{S}2}$ and $\gamma_{1}=1.1\gamma_{2}$.}
\label{FIG4}
\end{figure}

To quantify the nonreciprocal transport of FiM bimerons, the speed difference $\Delta v_{i} = \left|v_{i}(+j)\right| - \left|v_{i}(-j)\right|$ is defined.  
In Figs.~\ref{FIG4}(b)-\ref{FIG4}(d), $\Delta v_{i}$ is calculated as a function of the current density $j$, the damping $\alpha$ and the ratio of $M_{\text{S}1}$ and $M_{\text{S}2}$.
According to the fitting results shown in Fig.~\ref{FIG4}(b), the relationship between $\Delta v_{x}$ and $j$ is well described by this function $\Delta v_{x} = k_{2}j^{2} + k_{4}j^{4}$, where we take $\boldsymbol{p} = \boldsymbol{-e}_{y}$ and $G = 0$. 
To understand the results shown in Fig.~\ref{FIG4}(b), let us return to Eq.~(\ref{eq:1}).
For $G = 0$, Eq.~(\ref{eq:1}) is simplified as $v_{x} = F_{x}/(\alpha L_{xx})$, where $L_{xx}$ and $F_{x}$ are related to the bimeron's spin structure so their values are affected by the applied currents.
As mentioned earlier, opposite currents have different effects on the FiM bimeron with an asymmetrical shape, so that the value of $\left| F_{x}/L_{xx} \right|$ for a positive current is different from that for a negative current.
Thus, the relation between $F_{x}/L_{xx}$ and $j$ must include even terms in addition to odd terms, so the speed $v_{x}$ is written as a general polynomial form, $v_{x}=\sum_{l=1}k_{l}j^{l}/2$ with coefficient $k_{l}$.
Considering the first two terms of such a polynomial, the fitting results almost match the numerical simulations [Fig.~\ref{FIG4}(a)] (if more high-order terms are considered, the gap between the fitting results and numerical simulations will be narrowed).  
From the above speed $v_{x}$, we obtain the speed difference that only contains even terms, $\Delta v_{x} = k_{2}j^{2} + k_{4}j^{4}+\cdots$ (the magnitude of $k_{2}$, $k_{4}$ and $\cdots$ is directly related to the strength of nonreciprocal transport).
In addition, taking different damping $\alpha$, $\Delta v_{x}$ is calculated and summarized in Fig.~\ref{FIG4}(c), showing that $\Delta v_{x}$ is inversely proportional to $\alpha$, as indicated by this equation $v_{x} = F_{x}/(\alpha L_{xx})$.
Moreover, by changing the value of $M_{\text{S}1}$, $\Delta v_{x}$ and $\Delta v_{y}$ for different $M_{\text{S}1}/M_{\text{S}2}$ are obtained, as shown in Fig.~\ref{FIG4}(d), where $\Delta v_{x}$ reaches its maximum value at the angular momentum compensation point ($M_{\text{S}1}/M_{\text{S}2}=1.1$), and $\Delta v_{x}$ and $\Delta v_{y}$ almost are symmetric and antisymmetric about $M_{\text{S}1}/M_{\text{S}2}=1.1$, respectively.

\textit{Generalization of nonreciprocal transport to magnetic skyrmions.}~In the above sections, we discussed the nonreciprocal transport of the FiM bimeron with a positive topological charge $Q$. Such a nonreciprocal transport is also observed for the FiM bimeron with a negative $Q$ (see Fig. S17 of Supplemental Material~\cite{Shen_SM}).
The results of nonreciprocal transport of FiM bimerons can be extended to other types of spin textures with broken symmetry.
For general topological solitons, e.g. skyrmions, they have a symmetrical structure so nonreciprocal transport does not appear, while the bimerons under investigation have intrinsic asymmetrical shape, resulting in the presence of nonreciprocal dynamics.
Therefore, in order to attain the nonreciprocal transport, a break in structural symmetry of spin textures is required.
As shown in Fig. S18 of Supplemental Material~\cite{Shen_SM}, when an in-plane magnetic field is utilized to break the rotational symmetry of a FM skyrmion, the skyrmion driven by opposite currents exhibits nonreciprocal transport. 
Compared to the intrinsic asymmetry of bimerons ($k_{2} \approx -0.3$, Fig.~\ref{FIG4}), this externally induced asymmetry of skyrmions gives rise to a weak nonreciprocity ($k_{2} \approx -0.01 $, Fig. S18 of Supplemental Material~\cite{Shen_SM}).

\textit{Conclusions.}
We have analytically and numerically studied the drift and nonreciprocal motions of FiM bimerons driven by spin currents.
Our results demonstrate that due to the deformation of the bimeron's spin structure, spin currents may induce a drift speed which is perpendicular to the desired motion direction, even if the FiM system is at the angular momentum compensation point. 
Moreover, the symmetry analysis shows that since the FiM bimeron studied here has an asymmetrical shape, the bimeron driven by opposite currents exhibits nonreciprocal transport.
Our analysis of nonreciprocal transport of FiM bimerons is applicable to other types of spin textures with broken symmetry and our results are useful for building bimeron-based spintronic devices, such as bimeron diode and AC racetrack memory.

This study is supported by Guangdong Special Support Project (Grant No. 2019BT02X030), Shenzhen Fundamental Research Fund (Grant No. JCYJ20210324120213037), Shenzhen Peacock Group Plan (Grant No. KQTD20180413181702403), Pearl River Recruitment Program of Talents (Grant No. 2017GC010293) and National Natural Science Foundation of China (Grant Nos. 11974298 and 61961136006).
J.X. acknowledges the support by the National Natural Science Foundation of China (Grant No. 12104327).
X.Li acknowledges the support by the Guangdong Basic and Applied Basic Research Foundation (Grant No. 2019A1515111110).
X.Z. was an International Research Fellow of Japan Society for the Promotion of Science (JSPS). X.Z. was supported by JSPS KAKENHI (Grant No. JP20F20363).
O.A.T. acknowledges the support by the Australian Research Council (Grant No. DP200101027), the Cooperative Research Project Program at the Research Institute of Electrical Communication, Tohoku University (Japan), and by the NCMAS 2021 grant.
Q.S. acknowledges funding support from the Shenzhen-Hong Kong-Macau Science and Technology Program (Category C, Grant No. SGDX2020110309460000), Research Grant Council-Early Career Scheme (Grant No. 26200520), and the Research Fund of Guangdong-Hong Kong-Macao Joint Laboratory for Intelligent Micro-Nano Optoelectronic Technology (Grant No. 2020B1212030010).
G.Z. acknowledges the support by the National Natural Science Foundation of China (Grant Nos. 51771127, 51571126, and 51772004), the Scientific Research Fund of Sichuan Provincial Education Department (Grant Nos. 18TD0010 and 16CZ0006).
X.Liu acknowledges the support by the Grants-in-Aid for Scientific Research from JSPS KAKENHI (Grant Nos. JP20F20363 and JP21H01364).
M.E. acknowledges the support by the Grants-in-Aid for Scientific Research from JSPS KAKENHI (Grant Nos. JP17K05490 and JP18H03676) and the support by CREST, JST (Grant Nos. JPMJCR16F1 and JPMJCR20T2).
%




\begin{thebibliography}{99}
%

\bibitem{Takashima_PRB2018} R.~Takashima, Y.~Shiomi, and Y.~Motome, Phys. Rev. B \textbf{98}, 020401(R) (2018).

\bibitem{Tokura_NC2018} Y.~Tokura and N.~Nagaosa, Nat. Commun. \textbf{9}, 3740 (2018).

\bibitem{Fan_NanoLett2019} Y.~Fan, Q.~Shao, L.~Pan, X.~Che, Q.~He, G.~Yin, C.~Zheng, G.~Yu, T.~Nie, M.~R.~Masir, A.~H.~MacDonald, and K.~L.~Wang, Nano Lett. \textbf{19}, 692 (2019).

\bibitem{Li_RMP2012} N.~Li, J.~Ren, L.~Wang, G.~Zhang, P.~H{\"a}nggi, and B.~Li, Rev. Mod. Phys. \textbf{84}, 1045 (2012).

\bibitem{Whyte_NC2015} J.~R.~Whyte and J.~M.~Gregg, Nat. Commun. \textbf{6}, 7361 (2015).

\bibitem{Xing_JAP2016} X.~Xing, P.~W.~T.~Pong, and Y.~Zhou, J. Appl. Phys. \textbf{120}, 203903 (2016).

\bibitem{Song_JMMM2021} L.~Song, H.~Yang, B.~Liu, H.~Meng, Y.~Cao, and P.~Yan, J. Magn. Magn. Mater. \textbf{532}, 167975 (2021).

\bibitem{Franken_NatNano2012} J.~H.~Franken, H.~J.~M.~Swagten, and B.~Koopmans, Nat. Nanotechnol. \textbf{7}, 499 (2012).

\bibitem{Ideue_NatPhys2017} T.~Ideue, K.~Hamamoto, S.~Koshikawa, M.~Ezawa, S.~Shimizu, Y.~Kaneko, Y.~Tokura, N.~Nagaosa, and Y.~ Iwasa, Nat. Phys. \textbf{13}, 578 (2017).

\bibitem{Toyoda_PRL2015} S.~Toyoda, N.~Abe, S.~Kimura, Y.~H.~Matsuda, T.~Nomura, A.~Ikeda, S.~Takeyama, and T.~Arima, Phys. Rev. Lett. \textbf{115}, 267207 (2015).

\bibitem{Tateno_PRAppl2020} S.~Tateno and Y.~Nozaki, Phys. Rev. Appl. \textbf{13}, 034074 (2020).

\bibitem{Roszler_NATURE2006} U.~K. R{\"o}{\ss}ler, A.~N. Bogdanov, and C. Pfleiderer, Nature \textbf{442}, 797 (2006).

\bibitem{Jung_PRB2021} D.-H.~Jung, H.-S.~Han, N.~Kim, G.~Kim, S.~Jeong, S.~Lee, M.~Kang, M.-Y.~Im, and K.-S.~Lee, Phys. Rev. B \textbf{104}, L060408 (2021).

\bibitem{Zhao_Nanoscale2020} L.~Zhao, X.~Liang, J.~Xia, G.~Zhao, and Y.~Zhou, Nanoscale \textbf{12}, 9507 (2020).

\bibitem{Fook_SciRep2016} H.~Fook, W.~Gan, and W.~Lew, Sci. Rep. \textbf{6}, 21099 (2016).

\bibitem{Wang_APL2020} J.~Wang, J.~Xia, X.~Zhang, X.~Zheng, G.~Li, L.~Chen, Y.~Zhou, J.~Wu, H.~Yin, R.~Chantrell, and Y.~Xu, Appl. Phys. Lett. \textbf{117}, 202401 (2020).

\bibitem{Parkin_Sci2008} S.~S. Parkin, M. Hayashi, and L. Thomas, Science \textbf{320}, 190 (2008).

\bibitem{Gomonay_PRL2016} O. Gomonay, T. Jungwirth, and J. Sinova, Phys. Rev. Lett. \textbf{117}, 017202 (2016).

\bibitem{Muhlbauer_Sci2009} S. M{\"u}hlbauer, B. Binz, F. Jonietz, C. Pfleiderer, A. Rosch, A. Neubauer, R. Georgii, and P. B{\"o}ni, Science \textbf{323}, 915 (2009).

\bibitem{Gobel_PhysRep2021} B. G{\"o}bel, I. Mertig, and O.~A. Tretiakov, Phys. Rep. \textbf{895}, 1 (2021).

\bibitem{Ezawa_PRB2011} M. Ezawa, Phys. Rev. B \textbf{83}, 100408(R) (2011).

\bibitem{Lin_PRB2015} S.~Z. Lin, A. Saxena, and C.~D. Batista, Phys. Rev. B \textbf{91}, 224407 (2015). 

\bibitem{Leonov_PRB2017} A.~O. Leonov and I. K{\'e}zsm{\'a}rki, Phys. Rev. B \textbf{96}, 014423 (2017).

\bibitem{Kharkov_PRL2017} Y.~A. Kharkov, O.~P. Sushkov, and M. Mostovoy, Phys. Rev. Lett. \textbf{119}, 207201 (2017).

\bibitem{Yu_Nat2018} X.~Z. Yu, W. Koshibae, Y. Tokunaga, K. Shibata, Y. Taguchi, N. Nagaosa, and Y. Tokura, Nature \textbf{564}, 95 (2018).

\bibitem{Gobel_PRB2019} B. G{\"o}bel, A. Mook, J. Henk, I. Mertig, and O.~A. Tretiakov, Phys. Rev. B \textbf{99}, 060407(R) (2019).

\bibitem{Murooka2019} R. Murooka, A.~O. Leonov, K. Inoue, and J. Ohe, Sci. Rep. \textbf{10}, 396 (2020).

\bibitem{Shen_PRL2020} L. Shen, J. Xia, X. Zhang, M. Ezawa, O.~A. Tretiakov, X. Liu, G. Zhao, and Y. Zhou, Phys. Rev. Lett. \textbf{124}, 037202 (2020).

\bibitem{Zhang_Bimeron2020} X. Zhang, J. Xia, L. Shen, M. Ezawa, O.~A. Tretiakov, G. Zhao, X. Liu, and Y. Zhou, Phys. Rev. B \textbf{101}, 144435 (2020).

\bibitem{Li_npj2020} X. Li, L. Shen, Y. Bai, J. Wang, X. Zhang, J. Xia, M. Ezawa, O.~A. Tretiakov, X. Xu, M. Mruczkiewicz, M. Krawczyk, Y. Xu, F.~L. Evans, R.~W. Chantrell, and Y. Zhou, npj Comput. Mater. \textbf{6}, 169 (2020).

\bibitem{Nagase_NC2021} T.~Nagase, Y.-G.~So, H.~Yasui, T.~Ishida, H.~K.~Yoshida, Y.~Tanaka, K.~Saitoh, N.~Ikarashi, Y.~Kawaguchi, M.~Kuwahara, and M. Nagao, Nat. Commun. \textbf{12}, 3490 (2021).

\bibitem{Woo_NATCOM2018} S. Woo, K.~M. Song, X. Zhang, Y. Zhou, M. Ezawa, X. Liu, S. Finizio, J. Raabe, N.~J. Lee, S.~I. Kim, S.~Y. Park, Y. Kim, J.~Y. Kim, D. Lee, O. Lee, J.~W. Choi, B.~C. Min, H.~C. Koo, and J. Chang, Nat. Commun. \textbf{9}, 959 (2018).

\bibitem{Kim_NatMat2017} K.-J. Kim, S.~K. Kim, Y. Hirata, S.-H. Oh, T. Tono, D.-H. Kim, T. Okuno, W.~S. Ham, S. Kim , G. Go, Y. Tserkovnyak, A. Tsukamoto, T. Moriyama, K.-J. Lee, and T. Ono, Nat. Mater. \textbf{16}, 1187 (2017).

\bibitem{Caretta_NatNano2018} L.~Caretta, M.~Mann, F.~B{\"u}ttner, K.~Ueda, B.~Pfau, C.~M. G{\"u}nther, P. Hessing, A. Churikova, C. Klose, M. Schneider, D. Engel, C. Marcus, D. Bono, K. Bagschik, S. Eisebitt, and G.~S.~D. Beach, Nat. Nanotechnol. \textbf{13}, 1154 (2018).

\bibitem{Xu_PRM2021} T. Xu, Z. Chen, H.-A. Zhou, Z. Wang, Y. Dong, L. Aballe, M. Foerster, P. Gargiani, M. Valvidares, D.~M. Bracher, T. Savchenko, A. Kleibert, R. Tomasello, G. Finocchio, S.-G. Je, M.-Y. Im, D.~A. Muller, and W. Jiang, Phys. Rev. Mater. \textbf{5}, 084406 (2021).

\bibitem{Gao_Nature2020} S. Gao, H. Diego Rosales, F.~A. G{\'o}mez Albarrac{\'i}n, V. Tsurkan, G. Kaur, T. Fennell, P. Steffens, M. Boehm, P. {\v{C}}erm{\'a}k, A. Schneidewind, E. Ressouche, D.~C. Cabra, C. R{\"u}egg, and O. Zaharko, Nature \textbf{586}, 37 (2020).

\bibitem{Jani_Nature2021} H. Jani, J.-C. Lin, J. Chen, J. Harrison, F. Maccherozzi, J. Schad, S. Prakash, C.-B. Eom, A. Ariando, T. Venkatesan, and P.~G. Radaelli, Nature \textbf{590}, 74 (2021).

\bibitem{Nagaosa_NNANO2013} N. Nagaosa and Y. Tokura, Nat. Nanotechnol. \textbf{8}, 899 (2013).

\bibitem{Finocchio_JPD2016} G. Finocchio, F. B{\"u}ttner, R. Tomasello, M. Carpentieri, and M. Kl{\"a}ui, J. Phys. D: Appl. Phys. \textbf{49}, 423001 (2016).

\bibitem{Fert_NATREVMAT2017} A. Fert, N. Reyren, and V. Cros, Nat. Rev. Mat. \textbf{2}, 17031 (2017).

\bibitem{ES_JAP2018} K. Everschor-Sitte, J. Masell, R. M. Reeve, and M. Kl{\"a}ui, J. Appl. Phys. \textbf{124}, 240901 (2018).

\bibitem{Zhou_NSR2018} Y. Zhou, Natl. Sci. Rev. \textbf{6}, 210 (2019).

\bibitem{Zhang_JPCM2020} X. Zhang, Y. Zhou, K. M. Song, T.-E. Park, J. Xia, M. Ezawa, X. Liu, W. Zhao, G. Zhao, and S. Woo,
J. Phys. Condens. Matter \textbf{32}, 143001 (2020).

\bibitem{Sampaio_NatNano2013} J. Sampaio, V. Cros, S. Rohart, A. Thiaville, and A. Fert, Nat. Nanotechnol. \textbf{8}, 839 (2013).

\bibitem{Jiang_NatPhys2017} W. Jiang, X. Zhang, G. Yu, W. Zhang, X. Wang, M. Benjamin Jungfleisch, John E. Pearson, X. Cheng, O. Heinonen, K. L. Wang, Y. Zhou, A. Hoffmann, and Suzanne G. E. te Velthuis, Nat. Phys. \textbf{13}, 162 (2017).

\bibitem{Litzius_NatPhys2017} K. Litzius, I. Lemesh, B. Kr{\"u}ger, P. Bassirian, L. Caretta, K. Richter, F. B{\"u}ttner, K. Sato, O. A. Tretiakov, J. F{\"o}rster, R. M. Reeve, M. Weigand, I. Bykova, H. Stoll, G. Sch{\"u}tz, G. S. D. Beach, and M. Kl{\"a}ui, Nat. Phys. \textbf{13}, 170 (2017).

\bibitem{Zang_PRL2011} J. Zang, M. Mostovoy, J.~H. Han, and N. Nagaosa, Phys. Rev. Lett. \textbf{107}, 136804 (2011).

\bibitem{Zhang_SREP2016A} X. Zhang, Y. Zhou, and M. Ezawa, Sci. Rep. \textbf{6}, 24795 (2016).

\bibitem{Barker_PRL2016} J. Barker and O.~A. Tretiakov, Phys. Rev. Lett. \textbf{116}, 147203 (2016).

\bibitem{Potkina_JAP2020} M.~N. Potkina, I.~S. Lobanov, H. J{\'o}nsson, and V.~M. Uzdin, J. Appl. Phys. \textbf{127}, 213906 (2020).

\bibitem{Hirata_NatNano2019} Y. Hirata, D.-H. Kim, S.~K. Kim, D.-K. Lee, S.-H. Oh, D.-Y. Kim, T. Nishimura, T. Okuno, Y. Futakawa, H. Yoshikawa, A. Tsukamoto, Y. Tserkovnyak, Y. Shiota, T. Moriyama, S.-B. Choe, K.-J. Lee, and T. Ono, Nat. Nanotechnol. \textbf{14}, 232 (2019).

\bibitem{Augustin_NC2021} M. Augustin, S. Jenkins, R.~F.~L. Evans, K.~S. Novoselov, and E.~J.~G. Santos, Nat. Commun. \textbf{12}, 185 (2021).

\bibitem{Lu_NC2020} X. Lu, R. Fei, L. Zhu, and L. Yang, Nat. Commun. \textbf{11}, 4724 (2020).

\bibitem{Sun_NC2020} W. Sun, W. Wang, H. Li, G. Zhang, D. Chen, J. Wang, and Z. Cheng, Nat. Commun. \textbf{11}, 5930 (2020). 

\bibitem{Shen_PRB2020} L. Shen, X. Li, J. Xia, L. Qiu, X. Zhang, O.~A. Tretiakov, M. Ezawa, and Y. Zhou, Phys. Rev. B \textbf{102}, 104427 (2020).

\bibitem{Zhang_APL2021} X. Zhang, J. Xia, M. Ezawa, O.~A. Tretiakov, H.~T. Diep, G. Zhao, X. Liu, and Y. Zhou, Appl. Phys. Lett. \textbf{118}, 052411 (2021).

\bibitem{Gilbert_IEEE2004} T.~L. Gilbert, IEEE Trans. Magn. \textbf{40}, 3443 (2004).

\bibitem{Unal_PRAppl2016} A.~A. {\"U}nal, S. Valencia, F. Radu, D. Marchenko, K.~J. Merazzo, M. V{\'a}zquez, and J. S{\'a}nchez-Barriga, Phys. Rev. Appl. \textbf{5}, 064007 (2016).

\bibitem{Leonov_JPCM2016} A.~O.~Leonov, T.~L.~Monchesky, J.~C. Loudon, and A.~N. Bogdanov, J. Phys. Condens. Matter \textbf{28}, 35LT01 (2016).

\bibitem{Hals_PRL2011} K.~M.~D. Hals, Y. Tserkovnyak, and A. Brataas, Phys. Rev. Lett. \textbf{106}, 107206 (2011).

\bibitem{Kim_PRB2017} S.~K. Kim, K.-J. Lee, and Y. Tserkovnyak, Phys. Rev. B \textbf{95}, 140404(R) (2017).

\bibitem{Shen_SM} See Supplemental Material at http:// for the details of numerical simulations and analytical derivations for the FiM bimerons, which cites Refs.~\cite{Gilbert_IEEE2004,Rohart_PRB2013,Tveten_PRB2016,Barker_PRL2016,Shen_PRB2020,Shen_PRL2020,Li_npj2020,Thiele_PRL1973,Baltz_RMD2018,Kim_PRB2017,Sampaio_NatNano2013,Chen_arXiv2021}.

\bibitem{Slonczewski_JMMM1996} J.~C. Slonczewski, J. Magn. Magn. Mater. \textbf{159}, L1 (1996).

\bibitem{Tomasello_SciRep2014} R. Tomasello, E. Martinez, R. Zivieri, L. Torres, M. Carpentieri, and G. Finocchio, Sci. Rep. \textbf{4}, 6784 (2014).

\bibitem{Komineas_PRB2015} S. Komineas and N. Papanicolaou, Phys. Rev. B \textbf{92}, 064412 (2015).

\bibitem{Thiele_PRL1973} A.~A. Thiele, Phys. Rev. Lett. \textbf{30}, 230 (1973).

\bibitem{Tveten_PRL2013} E.~G. Tveten, A. Qaiumzadeh, O.~A. Tretiakov, and A. Brataas, Phys. Rev. Lett. \textbf{110}, 127208 (2013).

\bibitem{Tretiakov_PRL2008} O.~A. Tretiakov, D. Clarke, G.~W. Chern, Y.~B. Bazaliy, and O. Tchernyshyov, Phys. Rev. Lett. \textbf{100}, 127204 (2008).

\bibitem{Reichardt_arXiv} C.~Reichhardt, C.~J.~O.~Reichhardt, and M.~V. Milo{\v{s}}evi{\'c}, arXiv:2102.10464 (2021).

\bibitem{Bellizotti_PRB2021} J.~C. Bellizotti Souza, N.~P.~Vizarim, C.~J.~O.~Reichhardt, C.~Reichhardt, and P.~A.~Venegas, Phy. Rev. B \textbf{104}, 054434 (2021).

\bibitem{Gobel_SciRep2021} B.~G{\"o}bel and I.~Mertig, Sci. Rep. \textbf{11}, 3020 (2021).

\bibitem{Ma_PRB2017} X. Ma, C.~J. Olson Reichhardt, and C. Reichhardt, Phys. Rev. B \textbf{95}, 104401 (2017).

\bibitem{Reichhardt_ARCMP2017} C.~J. Olson Reichhardt and C. Reichhardt, Annu. Rev. Condens. Matter Phys. \textbf{8}, 51 (2017).

\bibitem{Chen_arXiv2021} Z. Chen, X. Zhang, Y. Zhou, and Q. Shao, arXiv:2112.04073 (2021).

\bibitem{Rohart_PRB2013} S. Rohart and A. Thiaville, Phys. Rev. B \textbf{88}, 184422 (2013).

\bibitem{Tveten_PRB2016} E.~G. Tveten, T. M{\"u}ller, J. Linder, and A. Brataas, Phys. Rev. B \textbf{93}, 104408 (2016).

\bibitem{Baltz_RMD2018} V. Baltz, A. Manchon, M. Tsoi, T. Moriyama, T. Ono, and Y. Tserkovnyak, Rev. Mod. Phys. \textbf{90}, 015005 (2018).

\end{thebibliography}
\end{document}